\documentclass[preprint]{rmaa}

\usepackage{epstopdf}
\usepackage{textcomp}
\usepackage{multirow} 
\title{Radio Continuum Emission from FS CMa Stars}

 \author{
Luis F. Rodr\'\i guez\altaffilmark{1,2},
Alejandro B\'aez-Rubio\altaffilmark{3,1}
and
Anatoly S. Miroshnichenko\altaffilmark{4}}
  \altaffiltext{1}{Centro de Radioastronom\'{\i}a y Astrof\'{\i}sica, Universidad
Nacional Aut\'onoma de M\'exico, Campus Morelia} 

\altaffiltext{2}{Astronomy Department, Faculty of Science, King Abdulaziz University, P.O.
Box 80203, Jeddah 21589, Saudi Arabia}

\altaffiltext{3}{Centro de Astrobiolog\'\i a (CSIC/INTA)}
 \altaffiltext{4}{Department of Physics and Astronomy, University of North Carolina at Greensboro}

\fulladdresses{
\item Luis F. Rodr\'\i guez: Centro de Radiostronom\'{\i}a
y Astrof\'\i sica, 
UNAM, A. P. 3-72, 
(Xangari), 58089 Morelia, Michoac\'an, M\'exico
(l.rodriguez@crya.unam.mx).
\item Alejandro B\'aez-Rubio: Centro de Astrobiolog\'\i a (CSIC/INTA), 
Ctra. de Torrej\'on a Ajalvir km 4, 28850 Torrej\'on de Ardoz, Madrid, Spain 
(baezra@cab.inta-csic.es).
\item Anatoly S. Miroshnichenko: Department of Physics and Astronomy, University of North Carolina, 
Greensboro, NC 27402 USA 
(a\_mirosh@uncg.edu).
}

\shortauthor{Rodr\'\i guez et al.}
\shorttitle{Radio Continuum Emission from FS CMa Stars}

\SetVolume{50} \SetFirstPage{001} \SetYear{2011}
\resumen{
Las estrellas tipo FS CMa muestran espectro \'optico con l\'\i neas de
emisi\'on y fuertes excesos infrarrojos. Se sabe muy poco de sus caracter\'\i sticas
en ondas de radio. Hemos analizado datos de archivo del Very Large Array para buscar emisi\'on
de radiocontinuo en una muestra de ellas. Existen datos de buena calidad para
siete de las aproximadamente 40 estrellas FS CMa conocidas. De estas siete
estrellas, cinco resultan tener emisi\'on de radio asociada. Dos de estas
estrellas, CI Cam y MWC 300, han sido reportadas previamente en la literatura
como emisoras de radio. Presentamos y discutimos brevemente la detecci\'on de las
otras tres fuentes: FS CMa (el prototipo de la clase), AS 381, y MWC 922.
La emisi\'on de radio es muy probablemente libre-libre pero observaciones adicionales
se requieren para caracterizarla mejor.
}
 
\abstract{The FS CMa stars 
exhibit bright optical emission-line spectra and strong IR
excesses. Very little is known of their radio characteristics.
We analyzed archive Very Large Array data to search for radio continuum
emission in a sample of them. There are good quality data
for seven of the $\sim$40 known FS CMa stars. 
Of these seven stars, five turn out to have associated radio emission.
Two of these stars, CI Cam and MWC 300, have been previously reported
in the literature
as radio emitters. We present and briefly discuss the radio detection of
the other
three sources: FS CMa (the prototype of the class), AS 381, and MWC 922. 
The radio emission is most probably of a free-free nature but additional
observations are required to better characterize it.
}
      

\begin{document}

\maketitle

\section{Introduction}

The B[e] phenomenon is associated with stars at different evolutionary stages, going from 
the pre-main sequence to the planetary nebula stage. This phenomenon is
characterized by the simultaneous presence of low-excitation forbidden line emission and
strong infrared excess in the spectra of early-type stars.  
The group of B[e] stars includes
high- and low-mass evolved stars, intermediate-mass pre-main
sequence stars and symbiotic objects.
In more than 50\% of the confirmed B[e] stars the evolutionary stage is still unknown. 
These objects are generally called unclassified B[e] stars (e.g. Miroshnichenko 2007;
Borges Fernandes 2010).
This lack of a classification is mostly caused by the limited
knowledge regarding
their physical parameters, in particular the distance,
and the geometry of their circumstellar matter.

Recently, Miroshnichenko (2007) has noted that most of the unclassified B[e]
stars have unique observational properties that distinguish them with respect
to the rest of the B[e] stars and classified them within a new group: the FS CMa
stars. Miroshnichenko (2007) has proposed that this group of stars could be binary
systems that are currently undergoing or have recently undergone a phase of
rapid mass exchange associated with strong mass loss stored in a
circumbinary envelope. This scenario explains the higher IR excesses due to
circumstellar dust
despite its lower mass-loss rate with respect to sgB[e] stars and the fact that
FS CMa stars are not found in star forming regions.

As noted before, the determination of the geometry of the surrounding matter
via high-angular resolution observations
could provide valuable information on the nature of the unclassified
B[e] stars. We 
have started a program to search for FS CMa stars with detectable radio continuum emission
in unpublished archive data from the Very
Large Array (VLA)
of the NRAO\footnote{The National Radio
Astronomy Observatory is operated by Associated Universities
Inc. under cooperative agreement with the National Science Foundation.}.
The sources detected will
be observed in the future to obtain high quality images with subarcsecond angular resolution
using the ultrasensitive Expanded Very Large Array (EVLA) at centimeter
wavelengths and the Plateau de Bure interferometer (PdBI) at millimeter wavelengths.

\begin{table*}[htbp]
\footnotesize
  \setlength{\tabnotewidth}{1.0\columnwidth} % width of footnotes
    \tablecols{9} % number of columns they have to span
\small
  \caption{FS CMa Stars with Good Quality VLA Archive Observations}
    \begin{center}
	\begin{tabular}{lcccccccc}\hline\hline
	&\multicolumn{2}{c}{Position$^a$} & Flux & Wavelength & VLA & & Epoch of  \\
	\cline{2-3} 
	Star &  $\alpha$(J2000) & $\delta$(J2000) & Density (mJy) & (cm) & Conf. &
	Project & Observation \\ 
	\hline
	FS CMa & 06 28 17.39 &  $-$13 03 10.9 & 4.2$\pm$0.4 & 1.3 & DnC & AM570 & 1997 Sep 26 \\
	MWC 819 & 06 44 37.67 & +01 19 32.5 & $\leq$0.31$^b$ & 6.0 & B &  AP116 & 1986 Jul 29 \\
	MWC 922 & 18 21 16.06 &  $-$13 01 25.7 & 10.8$\pm$0.4 & 3.6 & B & AL329 & 1994 Jul 09 \\
	AS 381 & 20 06 39.86 & +33 14 30.0 & 3.3$\pm$0.7 & 20.0 & C & AW271 & 1990 Nov 14  \\
	V669 Cep & 22 26 38.71 & +61 13 31.6 & $\leq$0.57$^b$ & 20.0 & A & AM590 & 1998 Apr 04 \\
	\hline\hline
\tabnotetext{a}{The position of MWC 819 is from Cutri et al. (2003) and the position of V669 Cep
is from Hog et al. (2000). The positions of the other three stars are
from the VLA images presented here.}
\tabnotetext{b}{Three-sigma upper limit.}
  \label{tab:1}
\end{tabular}
\end{center}
\end{table*}

\begin{figure*}
\centering
\includegraphics[scale=0.5, angle=0]{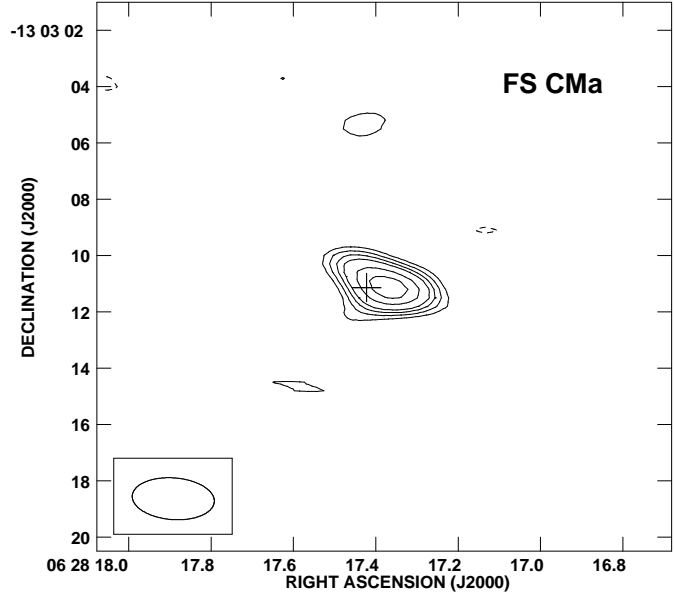}
 \caption{VLA contour image of the 1.3-cm continuum emission toward
FS CMa. Contours are -3, 3, 4, 5, 6, 8, and 10
times 0.36 mJy, the rms noise of the image. 
The synthesized beam, shown in the bottom left corner,
has half power full width dimensions of
$2\rlap.{''}92 \times 1\rlap.{''}48$, 
with the major axis at a position angle of $+86^\circ$. 
The cross marks the optical position of FS CMa from
Perryman et al. (1997).
}
  \label{fig1}
\end{figure*}

\begin{figure*}
\centering
\includegraphics[scale=0.5, angle=0]{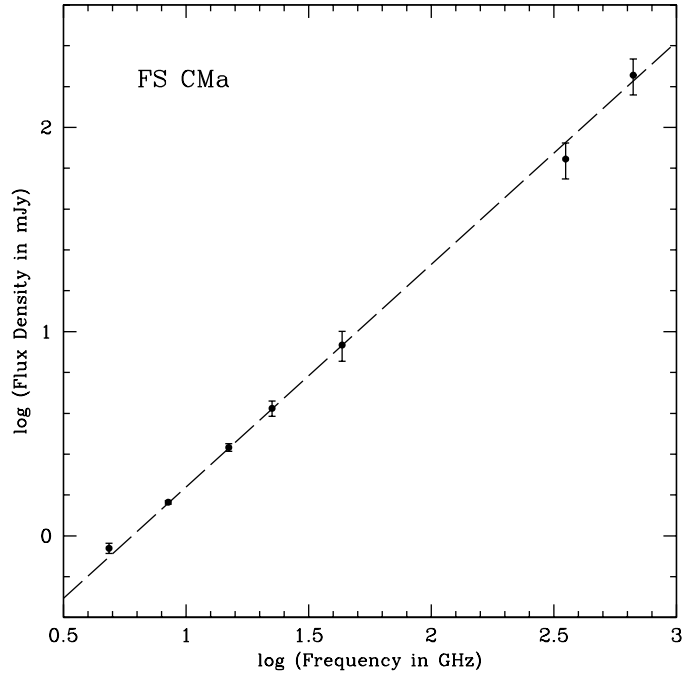}
 \caption{Radio spectrum of the star FS CMa. The five lower
 frequency data points are reported here while the two
 higher frequency data points are from Di Francesco et al. (2008).
 The dashed line marks the least-squares best fit to the data, given by
 $(S_\nu / mJy) = 0.141\pm 0.020 (\nu / GHz)^{1.09 \pm 0.03}$.
 }
   \label{fig2}
\end{figure*}

\begin{figure*}
\centering
\includegraphics[scale=0.5, angle=0]{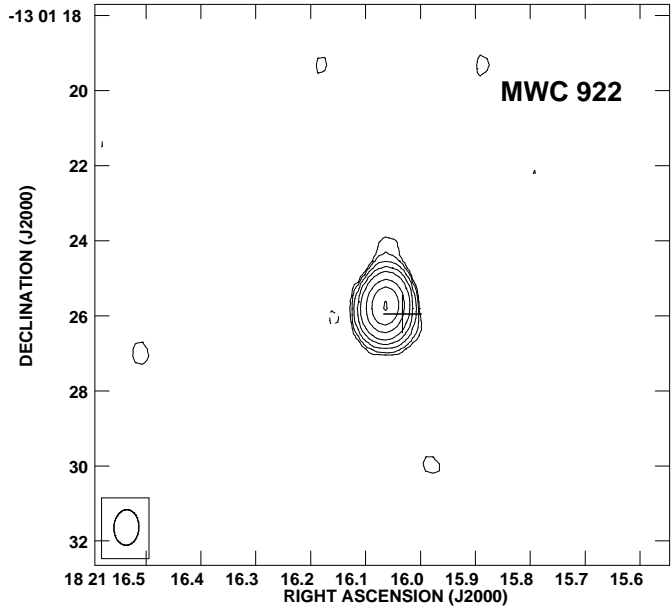}
 \caption{VLA contour image of the 3.6-cm continuum emission toward
MWC 922. Contours are -3, 3, 5, 10, 20, 40, 100, 200, and 400
 times 24 $\mu$Jy, the rms noise of the image.
 The synthesized beam, shown in the bottom left corner,
 has half power full width dimensions of
 $0\rlap.{''}95 \times 0\rlap.{''}66$,
 with the major axis at a position angle of $-3^\circ$.
 The cross marks the position of MWC 922 derived
 from the average of the 2MASS H, J and K images.
 }
   \label{fig3}
   \end{figure*}

\begin{figure*}
\centering
\includegraphics[scale=0.7, angle=0]{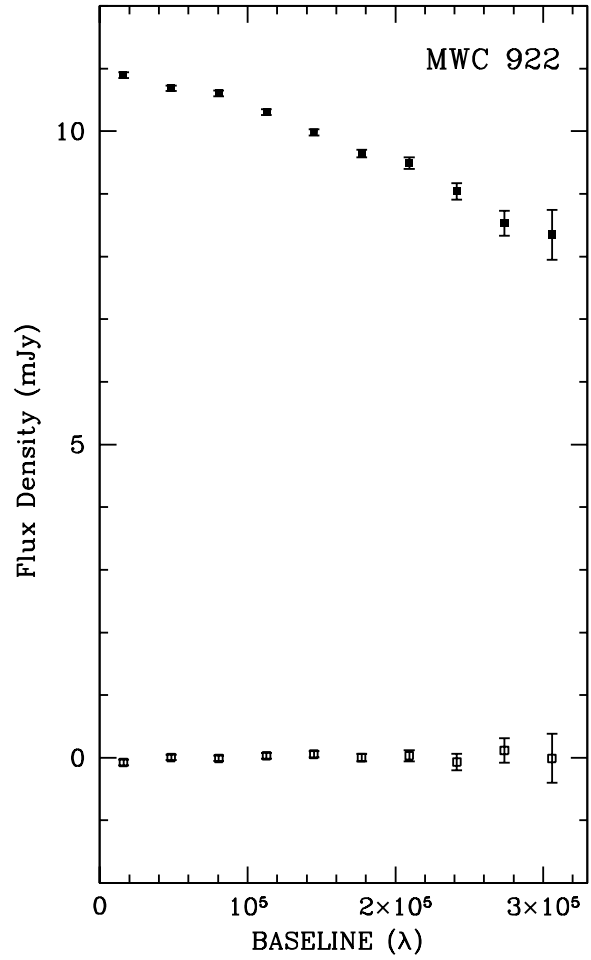}
 \caption{Real (filled squares) and imaginary (empty squares) components (given
 in mJy) of the
 emission from MWC 922 at 8.46 GHz as a function of baseline (given in wavelengths).
 The real component decreases with increasing baseline, indicating that the
 source is slightly resolved in these observations. 
 The imaginary component is consistent with zero, indicating that
 the source is symmetric about the phase center (the
 origin of the visibility plane) and has no significant structure on these spatial
 scales.
	 }
   \label{fig4}
\end{figure*}

\begin{table}[htbp]
\small
  \setlength{\tabnotewidth}{0.8\columnwidth} % width of footnotes
    \tablecols{3} % number of columns they have to span 
      \caption{Flux densities of 0607-085 and FS CMa for 1997 September 26}
	\begin{center}
	    \begin{tabular}{ccc}\hline\hline
	      Frequency &  Flux Density   & Flux Density \\
	       (GHz)  & 0607-085(Jy)$^a$ & FS CMa(mJy) \\
	       \hline
	       4.86  & 2.424$\pm$0.004  &  0.87$\pm$0.05 \\ 
		8.46 & 2.240$\pm$0.008  & 1.46$\pm$0.03 \\
		14.94 & 2.051$\pm$0.024 & 2.71$\pm$0.12 \\
		22.46 & 1.776$\pm$0.036 & 4.21$\pm$0.36  \\
		43.34 & 2.551$\pm$0.210  & 8.60$\pm$1.43 \\ 
	       \hline\hline
	       \tabnotetext{a}{The phase calibrator for all observations was 0607-085.}
   \label{tab2}
      \end{tabular}
 \end{center}
\end{table}

\begin{figure*}
\centering
\includegraphics[scale=0.5, angle=0]{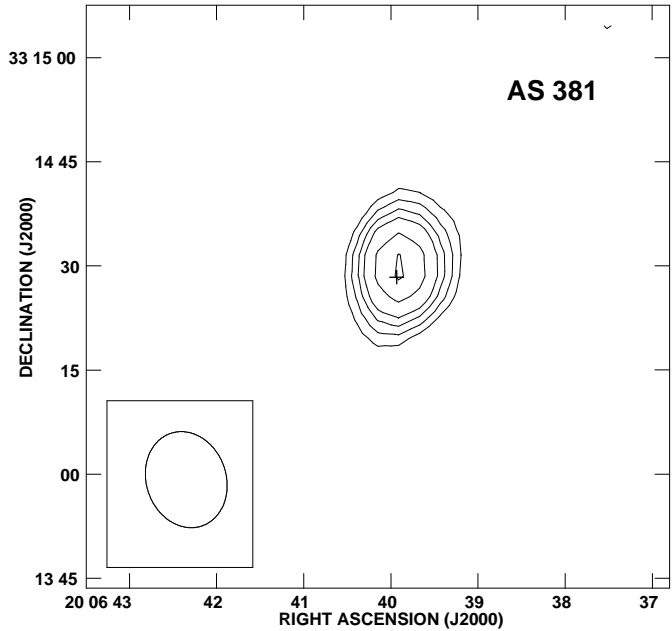}
 \caption{VLA contour image of the 20-cm continuum emission toward
AS 381. Contours are -3, 3, 4, 5, 6, 8, and 10
  times 0.32 mJy, the rms noise of the image.
  The synthesized beam, shown in the bottom left corner,
  has half power full width dimensions of
  $14\rlap.{''}1 \times 11\rlap.{''}5$,
  with the major axis at a position angle of $+19^\circ$.
  The cross marks the position of AS 381 derived
from the average of the 2MASS H, J and K images.
 }
    \label{fig5}
    \end{figure*}

\section{Data Reduction}

Of the $\sim$40 FS CMa stars and candidate stars reported
by Miroshnichenko (2007), only seven have good quality VLA observations 
(that could
provide a noise of order a few tenths of a mJy and a possible detection
at the mJy level). Of these seven stars,
CI Cam and MWC 300 have been previously reported as radio sources.
CI Cam is a high-mass X-ray binary that has been observed extensively with the
VLA, in particular after its 1998 outburst (e.g. Mioduszewski \& Rupen 2004).
MWC 300 has been observed in one occasion (1990 Feb 11) by Skinner et al. (1993), who
detected it with a flux density of 0.49$\pm$0.03 mJy at 3.6 cm. 

The remaining five stars are listed in Table 1, with the
parameters of their archive VLA observations.
The archive data from the Very
Large Array (VLA)
of the NRAO were edited and calibrated using the software package Astronomical Image
Processing System (AIPS) of NRAO. 

\section{Discussion on Individual Sources}

For the stars MWC 819 and V669 Cep only upper limits were obtained. However,
the other three stars, FS CMa, MWC 922, and AS 381 have associated radio continuum
emission and we discuss them in what follows.

\subsection{FS CMa}

This star is the prototype of the class and we find it is associated with a radio
continuum source (see Fig. 1). This source was observed at 6.0, 3.6, 2.0, 1.3
and 0.7 cm in the same observing session (1997 September 26)
and the observed flux densities are given in Table 2.
Di Francesco et al. (2008) report flux densities for FS CMa  
of 0.070$\pm$0.014 and 0.180$\pm$0.036 Jy at
850 and 450 $\mu$m, respectively.
These seven data points allow the analysis of its spectrum as shown in Figure 2.
Remarkably, the spectrum is well described over two decades of frequency
by a power-law of the form

$$\biggl[{{S_\nu} \over {mJy}} \biggr] = (0.141\pm 0.020) \biggl[{{\nu}
\over {GHz}} \biggr]^{1.09 \pm 0.03}.$$

This spectral index is consistent with partially optically-thick free-free emission.
It is significantly steeper than the spectrum from ionized winds
expanding at constant velocity, that
behaves as $S_\nu \propto \nu^{0.6}$ (i. e. Panagia \& Felli 1975).
The Herbig B[e] star MWC 297
(Cidale et al. 2000) also shows a spectral index of $\sim$1 from the
radio to the sub-mm (Sandell et al. 2011).
A spectral index of the order of 1 is also frequently found in
hypercompact H II regions (Ignace \& Churchwell 2004). This departure from
the expected value of 0.6 most probably indicates that the outflow
has velocity, temperature, or ionization fraction gradients with radius.
For example, assuming constant velocity and electron temperature in the outflow and
following Olnon (1975) and Panagia \& Felli (1975), the observed spectral
index of $\sim$1.1 implies an electron density gradient of
$n_e \propto r^{-2.8}$, steeper than the gradient of
$n_e \propto r^{-2.0}$ expected for a constant ionization fraction.

\subsection{MWC 922}

The radio source associated with this star (Fig. 3) is quite bright
at 3.6 cm. Unfortunately, there are
no observations at other frequencies that could allow the determination of
its spectral index.
The source is angularly resolved, as can be seen in the behavior of its
amplitude as a function of baseline (Fig. 4).
Analysis of the source with the task JMFIT of AIPS gives deconvolved
dimensions of $0\rlap.{''}28 \pm 0\rlap.{''}0.01 \times 0\rlap.{''}20 \pm 0\rlap.{''}0.01$
with a position angle of $169^\circ \pm 7^\circ$. From the measured flux density and
these angular dimensions, we obtain a brightness temperature of $\sim 3.3 \times 10^4$ K,
suggestive of partially optically thick free-free emission from photoionized gas.

Recently, Tuthill \& Lloyd (2007) reported detection of a biconical "Red
Square" nebula surrounding MWC 922 in the near-infrared.
This nebula extends for about 5$''$.
The radio continuum emission probably traces the very inner part of this structure. 

\subsection{AS 381}

AS 381 is a binary system with a spectrum that indicates 
the presence of both a hot (early B-type) star and a cool (K-type) star
(Miroshnichenko et al. 2002).
Of the three stars with radio continuum reported here, this is the only whose membership
is under debate since it has also been proposed as a possible galactic supergiant
candidate (sgB[e]; Miroshnichenko 2007).
%this is the only one that is
%a galactic supergiant candidate (sgB[e]; Miroshnichenko 2007). However, this classification
%is marginal because AS 381 has a luminosity
%of $10^{4.9 \pm 0.2}~L_\odot$, that overlaps within the error with the lower limit
%for sgB[e] stars of $10^{4.7}~L_\odot$. 
The characteristics of AS 381
allow to classify it as an evolved object with an initial mass of about 20
solar masses (Miroshnichenko et al. 2002).
%, while FS CMa stars can have luminosities as large as 
%$10^{4.5}~L_\odot$ (e.g. CI Cam has a
%luminosity of $L=10^{4.5/pm0.5} ~L_\odot$; Miroshichenko 2007).

This source is angularly unresolved, but given the modest angular resolution of
the observations ($\sim13''$) this does not provide additional information.

\section{Conclusions}

Using VLA archive data, we report the detection of radio continuum emission
toward three FS CMa stars: FS CMa, MWC 922, and AS 381.
Given that we only found good quality data for five stars, these results
suggest that detectable radio continuum emission could be common in
FS CMa stars. 

In the case of FS CMa, we combined the VLA data with
JCMT/SCUBA observations to show that its radio continuum spectrum
is well described by a single power law over two decades in
frequency.

Although the data do not have sufficient frequency coverage and
angular resolution to provide important new insight 
into this type of stars, the flux densities detected are relatively bright and
will allow in the future a detailed radio study of the spectrum and morphology of the sources.

%\adjustfinalcols

\acknowledgments
We are thankful to an anonymous referee for valuable comments that
improved the paper.
LFR is thankful for the support
of DGAPA, UNAM, and of CONACyT (M\'exico).
This publication makes use of data products from the Two Micron All Sky Survey, which is 
a joint project of the University of Massachusetts and the Infrared Processing and 
Analysis Center/California Institute of Technology, funded by the National 
Aeronautics and Space Administration and the National Science Foundation.
This research has made use of the SIMBAD database, 
operated at CDS, Strasbourg, France.

\vskip0.5cm

%\vspace*{4cm}


\begin{thebibliography}

%\bibitem{1988iras....1.....B} Beichman, C.~A., 
%Neugebauer, G., Habing, H.~J., Clegg, P.~E., 
% \& Chester, T.~J.\ 1988, Infrared astronomical satellite (IRAS) 
%catalogs and atlases.~Volume 1: Explanatory supplement, 1  

\bibitem{2010RMxAC..38...98B} Borges Fernandes, M.\ 
2010, Revista Mexicana de Astronomia y Astrofisica Conference Series, 38, 
98 

\bibitem{2000ASPC..214...87C} Cidale, L., Zorec, J., 
\& Morrell, N.\ 2000, IAU Colloq.~175: The Be Phenomenon in Early-Type Stars, 214, 87 

\adjustfinalcols

\bibitem{2003tmc..book.....C} Cutri, R.~M., et al.\ 
2003, The IRSA 2MASS All-Sky Point Source Catalog, NASA/IPAC Infrared 
Science Archive, http://irsa.ipac.caltech.edu/applications/Gator/ 

\bibitem{2008ApJS..175..277D} Di Francesco, J., 
Johnstone, D., Kirk, H., MacKenzie, T., 
\& Ledwosinska, E.\ 2008, \apjs, 175, 277 

\bibitem{2000A&A...355L..27H} H{\o}g, E., et al.\ 2000, \aap, 355, L27 

\bibitem{2004ApJ...610..351I} Ignace, R., \& Churchwell, E.\ 2004, \apj, 610, 351 


\bibitem{2004ApJ...615..432M} Mioduszewski, A.~J., \& Rupen, M.~P.\ 2004, \apj, 615, 432 

\bibitem{2002A&A...383..171M} Miroshnichenko, A.~S., et al.\ 2002, \aap, 383, 171 

\bibitem{2007ApJ...667..497M} Miroshnichenko, A.~S.\ 
2007, \apj, 667, 497 

\bibitem{1975A&A....39..217O} Olnon, F.~M.\ 1975, \aap, 39, 217 

\bibitem{1975A&A....39....1P} Panagia, N., \& Felli, M.\ 1975, \aap, 39, 1 

\bibitem{1997A&A...323L..49P} Perryman, M.~A.~C., et al.\ 1997, \aap, 323, L49 

\bibitem{2011ApJ...727...26S} Sandell, G., Weintraub, 
D.~A., \& Hamidouche, M.\ 2011, \apj, 727, 26 

\bibitem{1993ApJS...87..217S} Skinner, S.~L., Brown, 
A., \& Stewart, R.~T.\ 1993, \apjs, 87, 217 

\bibitem{2007Sci...316..247T} Tuthill, P.~G., \& Lloyd, J.~P.\ 2007, Science, 316, 247 

%\bibitem{1989ApJ...340..265W} Wood, D.~O.~S., \& Churchwell, E.\ 1989, \apj, 340, 265 



\end{thebibliography}
\end{document}